\documentstyle[preprint,revtex]{aps}
\begin{document}
IU/NTC 92--34   \hfill                           November 1992\\
\begin{title}
\center{The Nuclear Response in Delta-Isobar Region}\\
in the ($^3\!$He,t) Reaction
\end{title}
\author{V.F.Dmitriev}
\begin{instit}
Nuclear Theory Center, Indiana University, Bloomington IN 47408\\
\center{and}\\
Budker Institute of Nuclear
Physics, Novosibirsk, 630090, Russia
\end{instit}
\begin{abstract}

The excitation of a $\Delta$-isobar in a finite nucleus in charge--exchange
($^3\!$He,t) reaction is discussed in terms of a nuclear response function.
The medium effects modifying a $\Delta$- and a pion propagation were
considered for a finite size nucleus. The Glauber approach has been used
for distortion of a $^3\!$He and a triton in the initial and the final
states. The effects determining the peak positions and its width are
discussed.
Large displacement width for the $\Delta$ - h excitations and
considerable contribution of coherent pion production were found
for the reaction on $^{12}$C.
\end{abstract}
\newpage
        \section{INTRODUCTION}

The experimental studies of  nuclear response in charge--exchange reactions
were extended in the  eighties  to high excitation energies where a first
nucleon resonance the $\Delta$-isobar can be excited \cite{elle83}. The
detailed studies were done for the ($^3\!$He,t) charge--exchange reaction
at different projectile energies \cite{able84} and different targets
\cite{cont86}. The properties of the $\Delta$ excited in a nucleus were
found different compared to the case of the reaction on a single nucleon.
The difference was both in the peak position and the width of the resonance
excited in a complex nucleus. The review of the observed phenomena can be
found in the recent paper \cite{gaard91}.

The appealing explanation of this phenomenon is related to medium effects,
namely, the excitation of a pionic nuclear mode \cite{chanf84},
\cite{dmit85}, \cite{udaga90}, \cite{delo91} although another explanation
has been proposed as well \cite{oset89}. In this picture the $\Delta$ in
nuclear matter does not exist as separate resonance but forms a collective
excitation consisting of pionic, $\Delta$ and nucleon degrees of freedom.

At first sight one should not expect sizeable medium effects for
($^3\!$He,t) reaction since the reaction takes place at the surface of the
target \cite{esbe85}. For inelastic reactions, however, the absorption is
smaller and one should use absorption factor different from that used for
elastic scattering \cite{dmit89} providing both the medium effects and the
magnitude of inelastic cross--section. Another important point is the
account of the finite target size in the response function. As it will be
shown below different $\Delta$-hole multipoles are peaked at different
energies so, part of the observed width can be attributed to this
displacement. Besides, on finite nucleus the process of coherent pion
production is possible. The process is absent in infinite nuclear matter
and it contributes to the shift of the peak position as well.

Here we present the results of the absolute cross--section calculations of
the $^{12}$C($^3\!$He,t) reaction at 0$^\circ$ and at the kinetic energy of
the $^3\!$He T$_{He}$=2 GeV. In the next section we discuss the model for
the reaction amplitude that is a driving force for the nuclear response. In
the section~\ref{resp} the pionic response function of a finite nucleus is
discussed and the cross--section is calculated in the section ~\ref{cross}.

        \section{Reaction Amplitude.}
 \label{ampl}

We shall start from the discussion of the models for the elementary
charge--exchange reaction p(p,n)$\Delta^{++}$. From the early sixties it
was shown this reaction can be described by OPE model \cite{ferr61}, at
least at low momentum transfer. This analysis has been extended for wide
range of the proton energies in \cite{wolf69}, and has been repeated with
some minor modifications in connection to the analysis of p($^3\!$He,t)
$\Delta^{++}$ reaction in \cite{dmit86}. As it was shown all existing data
in the region of low momentum transfer are well described by OPE with the
soft monopole $\pi$NN- and $\pi$N$\Delta$- form factors

\begin{equation}
            F(q^2) = \frac{\Lambda^2 - \mu^2}{\Lambda^2 - q^2}
\label{form}
\end{equation}
The parameter $\Lambda = 650$ MeV at low proton energy and slightly
decreases with the proton energy reflecting increase of the absorption
effects at high energy. With these soft form factors (\ref{form}) the main
contribution to the $\Delta$-production comes from the direct graph shown
in Fig.1. The exchange contribution is small for the p(p,n)$\Delta^{++}$
reaction.

In this model the amplitude is completely longitudinal with respect to the
momentum transfer. In the other model used for the description of the
$\Delta$- production at 800 MeV proton energy \cite{verv79} the transverse
part of the amplitude was described by $\rho$- exchange and hard form
factors with $\Lambda$ = 1.2 GeV and $\Lambda$ = 1.7 GeV were used for
$\pi$N$\Delta$- and $\rho$N$\Delta$- vertexes. The magnitude of the
cross--section and the momentum transfer dependence are reproduced in this
model due to cancellation between the direct and the exchange parts of the
amplitude. The cancellation is rather delicate and at higher proton energy
it can be broken resulting in wrong momentum transfer dependence
\cite{dmit86}.

At very high energy the situation is different. The $\pi$-exchange
contribution decreases as ${\it s}^{-2}$, where {\it s} is the center--of
--mass energy squared, while for the $\rho$-exchange the decrease is
slower. Its contribution falls down like ${\it s}^{-2+{\alpha}(0)}$ for
small momentum transfer where ${\alpha}({\it t})$ is the corresponding
Regge--trajectory. Thus, in the asymptotic region at high energy one should
expect the dominance of the $\rho$-exchange even at forward angles.  Below 20
GeV the cross--section for forward scattering follows the $1/{\it s}^2$ law
\cite{karm73} so, the contribution of $\rho$-exchange at intermediate
energies is believed to be small.

For the p$(^3\!$He,t)${\Delta}^{++}$ reaction the situation is similar to
the (p,n) case. The $\pi$-exchange with the soft form factors (\ref{form})
gives reasonable description of the absolute cross--section and the tritium
spectrum at forward angle for all existing data \cite{dmit86}.
Nevertheless, at the kinetic energy of $^3\!$He 2 GeV, which is close in
kinematics to 800 MeV (p,n) one can get good description using $\pi +
\rho$- exchanges as well \cite{desg92}. It would be very desirable to
extend the last analysis to higher $^3\!$He energies.
\section{Nuclear Matter Response to the Pionic Probe.}
        \subsection{($^3\!$He,t) Cross--Section in Plane Wave Approximation.}
It is convenient to start with the plane waves for both projectile and
ejectile in order to obtain an expression for the cross--section that can
be easily generalized to the distorted waves. In PWIA the cross--section
is proportional to the matrix element, shown in Fig.1, squared and summed
over final nuclear and $\Delta$ - states.
\begin{equation}
 T = \int d^3r \Gamma_{{\pi}Het}({\bf r}) \cdot G_0({\bf r}-{\bf r}^\prime)
 \cdot\Gamma_{\pi N \Delta}({\bf r}^\prime)d^3r^\prime
\label{amp}
\end{equation}
For plane waves $\Gamma \sim exp({\imath{\bf qr}})$, it gives for the
cross--section
\begin{equation}
{{d^2{\sigma}}\over{dE^{\prime}d\Omega}} = {{M^2_{He}}\over{4{\pi}^2}}
{{p^\prime}\over{p}}\overline{\left|\Gamma_{\pi Het}({\bf q})\right|^2}
\sum_{\Delta h} \delta (\omega - E_{\Delta h})n_h{\left|\Gamma_{\pi N
\Delta}({\bf q})\right|^2}\cdot \left|G_0(q)\right|^2   \label{dsec}
\end{equation}
The expression under the sum is just imaginary part of the pionic
self--energy in nuclear medium.
\begin{equation}
\Im{m}\Pi_{\Delta}(\omega,{\bf q},{\bf q}) = \pi \sum_{\Delta h}\delta(
\omega - E_{\Delta h})\left|\Gamma_{\pi N\Delta}\right|^2n_h
\label{self}
\end{equation}
Using the pionic self--energy (\ref{self}) we obtain the final expression for
the cross--section suitable for inclusion of medium effects.
\begin{equation}
{{d^2{\sigma}}\over{dE^{\prime}d\Omega}} = {{M^2_{He}}\over{4{\pi}^3}}
{{p^\prime}\over{p}}\overline{\left|\Gamma_{\pi Het}({\bf q})\right|^2}
G_0^\ast(q)\cdot \Im{m}\Pi_{\Delta}(\omega,{\bf q},{\bf q})\cdot G_0(q).
\label{fsec}
\end{equation}
        \subsection{Medium Effects in Nuclear Matter}
The main effect of nuclear medium is the renormalization of the pion
propagator by intermediate $\Delta$-hole loops giving the major
contribution to the pionic self--energy (\ref{self}) near the
$\Delta$-resonance. To take it into account one must change in (\ref{fsec})
the bare pion propagator $G_0(q)$ for the dressed one $G(\omega,{\bf q})$,
where $$G(\omega,{\bf q}) = {1\over{q^2 - \mu^2 - \Pi_{\Delta}(\omega,{\bf
q})}}.$$ Making this change we are going out of the scope of the impulse
approximation.

The imaginary part of the bare pion propagator is equal to zero for
negative $q^2$. Using it we obtain $G^\ast(\omega,{\bf q})\cdot
\Im{m}\Pi_{\Delta} (\omega,{\bf q})\cdot G(\omega,{\bf q}) = -
\Im{m}G(\omega,{\bf q}).$ With these changes the cross--section
(\ref{fsec}) becomes
\begin{equation}
{{d^2{\sigma}}\over{dE^{\prime}d\Omega}} = -{{M^2_{He}}\over{4{\pi}^3}}
{{p^\prime}\over{p}}\overline{\left|\Gamma_{\pi Het}({\bf q})\right|^2}
\cdot \Im{m}G(\omega,{\bf q}).
\label{res}
\end{equation}
It is clear from (\ref{res}) the pion propagator $G(\omega,{\bf q})$ in
nuclear medium is just the response function to a virtual pion probe. The
excitation created by a virtual pion is no more pure $\Delta$-hole but a
superposition of the $\Delta$-hole and pionic degrees of freedom, which is
usually called the pionic mode.

The unquenched $\Delta$-hole self--energy (\ref{self}) produces too much of
attraction giving unreasonably low excitation energy for the pionic mode.
In order to make the description more accurate several effects should be
taken into account. First of all, more correct $\pi$N- scattering amplitude
reproducing $s_{1\over2}$ and $p_{1\over2}$ partial waves should be used
since we are interested in the energies lower than the $\Delta$ in vacuum.
For this purpose one should add the Born diagrams with a nucleon
intermediate state, {\it u}-channel $\Delta$ diagram and a $\sigma$-term
arising from the $\sigma$-commutator \cite{olss75}. Second, the
short--range N$\Delta$- correlations must be taken into account.
\begin{equation}
 W({\bf r}_1,{\bf r}_2) = {f_{\Delta}^2\over{\mu^2}}g_{\Delta}'({\bf S}_1
 ^\dagger\cdot{\bf S}_2)({\bf T}_1^\dagger\cdot{\bf T}_2)\delta ({\bf r}_1
 - {\bf r}_2).
\label{corr}
\end{equation}
In nuclear matter the effect of short--range correlations can be accounted
in the following way. Let us define the $\Delta$-hole response function
$\chi (\omega,{\bf q})$ by
$$ \Pi_\Delta
(\omega,{\bf q}) ={\left(\frac{\Lambda^2-\mu^2}{\Lambda^2-q^2}\right)}^2
{\bf q}^2\cdot\chi (\omega,{\bf q}).$$ Then,
\begin{equation}
\tilde{\Pi}_{\Delta}(\omega,{\bf q}) = {\Pi_\Delta (\omega,{\bf q})\over1 -
g'_{\Delta}\cdot\chi (\omega,{\bf q})}.
\end{equation}
 Similar effect should be taken into account for the nucleon--hole response
 function as well. But, in the region of interest in excitation energy
 its contribution is negligible and will be omitted below. In contrast,
 the backward $\Delta$-hole loops and the $\sigma$-term must be retained
 since they have their own dependence on the virtual pion mass -- $q^2$
 that influences the position of the pionic branch in nuclear matter.

 Finally, the virtual pion can be absorbed in nuclear medium emitting two or
 more nucleons. In the resonance region the absorption goes via intermediate
 $\Delta$ and it can be described by a $\Delta$-nucleus optical potential.
 At lower energies the other mechanisms with different intermediate states
 contribute to the absorption.
  To take it into account we use
 the Ericson--Ericson optical potential from pionic atoms \cite{erer66}
 $$V_{2N} = - 4\pi\imath{ImC}\cdot{n^{2}(r)}\cdot{\bf q}^2,$$ where $n(r)$
 is the nuclear matter density. For pionic atoms $ImC =
 0.08\cdot\left({\hbar\over\mu{c}}\right)^6$.  It was obtained from the fit
 of the mesoatomic level width and contains all absorption mechanisms
 including the intermediate $\Delta$. Using this value in our case will
 give double counting since intermediate $\Delta$ is considered explicitly.
 Thus, $ImC$ should be considered as a free parameter accounting another
 absorption mechanisms.

 With these corrections the pion self--energy in nuclear matter is
 \begin{equation}
 \Pi(\omega,{\bf q}) = \tilde{\Pi}_{\Delta}(\omega,{\bf q}) + {1\over{f_{\pi}
 ^2}}\left(1 - {2q^2\over\mu^2}\right)\sigma(0)\cdot{n(r)} + V_{2N},
\label{full}
\end{equation}
where $f_\pi$ is the pion decay constant $f_\pi = 133$ MeV and $\sigma{(0)}$
is the $\sigma$-commutator for forward pion scattering.
$\tilde{\Pi}_{\Delta}(\omega,{\bf q})$ includes both forward and backward
$\Delta$-hole loops corrected for short--range correlations.
        \section{Response Function of a Finite Nucleus.}
\label{resp}
For a finite nucleus it is convenient to work in the configuration space
where the self--energy (\ref{full}) and the pion propagator become the
functions of two distinct variables instead of functions of the distance
between coordinate points in nuclear matter. $$\Pi(\omega,{\bf r} - {\bf r'}
) \rightarrow \Pi(\omega,{\bf r},{\bf r'});$$ $$V_{2N} = 4\pi\imath
{ImC}({\bf\nabla}\cdot{n^2(r)}\cdot{\bf\nabla})\delta({\bf r} - {\bf r'}).$$
 The $\pi N\Delta$ - vertex in the configuration space is
\begin{equation}
 \Gamma_{\pi{N}\Delta}({\bf r},{\bf r'}) = -\imath{\bf T}({\bf S}\cdot
{\bf\nabla}){f_\Delta\over\mu}{{\Lambda^2 - \mu^2}\over{4\pi}}
 {\exp(-\kappa\left|{\bf r} - {\bf r'}\right|)\over\left|{\bf r} - {\bf r'}
 \right|},
\label{pndel}
\end{equation}
 where $\kappa^2= \Lambda^2-\omega^2$.
        \subsection{Multipole Expansion}
 For a spherical nucleus the self--energy (\ref{full}) has simple multipole
 expansion $$ \Pi(\omega,{\bf r},{\bf r'}) = \sum_{JM}\Pi_J(r,r')
 Y_{JM}^\ast({\bf n})Y_{JM}({\bf n'}).$$ The similar expansion exists for
 the $\pi{N}\Delta$-vertex
\begin{equation}
 \Gamma_{\pi{N}\Delta}({\bf r}-{\bf r'})=
 \sum_{JLM}\Gamma_{JL}^{0}(r,r')Y_{JM}^\ast({\bf n})T_{JM}^{L}({\bf n'}),
\end{equation}
 where the tensor operator
\begin{equation}
   T_{JM}^{L}({\bf n'}) = {\bf S}\cdot{\bf Y}_{JM}^{L}({\bf n})
 = [{\bf S}\wedge{Y_{Lm}({\bf n})}]_{JM},
\label{tens}
\end{equation}
 and the radial
 vertex $\Gamma_{JL}^{0}(r,r')$ is
\begin{equation}
 \Gamma_{JL}^0(r,r') = \imath(L-J){f_\Delta\over\mu}{{2\kappa^{2}(
 \Lambda^2-\mu^2)}\over{\pi}}
 \sqrt{\frac{J+L+1}{2(2J+1)}}\left\{\begin{array}{l}
 i_{L}(\kappa r)k_{J}(\kappa r') \mbox{    if $r<r',$} \\
 -i_{J}(\kappa r')k_{L}(\kappa r) \mbox{    if $r>r',$}
\end{array}\right.
\label{gamd0}
\end{equation}
 here $i_{L}(x)$ and $k_{L}(x)$ are the spherical Bessel functions with
 an imaginary argument. This is the advantage of using the simple scalar
 formfactors~(\ref{form}).

 The $\Delta$-hole response function $\chi$ can be expanded using the set
 of tensor operators~(\ref{tens})
\begin{eqnarray}
\nonumber
\chi^{J}_{LL'}(r,r')&=&{1\over{2J+1}}\sum_{j_{N}j_{\Delta}l_{N}l_{\Delta}}
\langle j_{N}l_{N}\|T^{L}_{J} \|j_{\Delta}l_{\Delta}\rangle\langle
j_{\Delta}l_{\Delta}\|T^{L'}_{J}\| j_{N}l_{N}\rangle[g_{j_{\Delta}
l_{\Delta}}(\omega+\epsilon_{j_{N}l_{N}};r,r') \\
 &+&g_{j_{\Delta}l_{\Delta}}
(-\omega+\epsilon_{j_{N}l_{N}};r,r')]\cdot n_{j_{N}l_{N}}\cdot
R_{j_{N}l_{N}}(r)R_{j_{N}l_{N}}(r'),
\label{fresp}
\end{eqnarray}
 where $n_{j_{N}l_{N}}$ are the nucleon occupation numbers,
 $R_{j_{N}l_{N}}(r)$ the radial wave function of bounded nucleon and the
 $\epsilon_{j_{N}l_{N}}$ is its energy.
$g_{j_{\Delta}l_{\Delta}}(\omega;r,r')$ is the Green function of the
radial Schr\"odinger
equation for the $\Delta$ moving in the mean nuclear optical potential.
It was calculated using two independent solutions of the radial
Schr\"odinger equation.

The $\Delta$-hole contribution to the pion self--energy was calculated
using the following expression
\begin{equation}
\tilde{\Pi}_{\Delta}^{J}(\omega;r,r')=\sum_{LL'}\int \rho^{2}d\rho{\rho'}
^{2} d\rho'
\Gamma^0_{JL}(r,\rho)\chi^{J}_{LL'}(\rho,\rho')\Gamma_{JL'}^{\ast}(\rho,r');
\end{equation}
where $\Gamma_{JL}$ related to $\Gamma^{0}_{JL}$ via linear integral
equation accounting the short--range correlations (\ref{corr})
\begin{equation}
\Gamma_{JL}(r,\rho)=\Gamma^{0}_{JL}(r,\rho) +g'\left({f_{\Delta}\over\mu}
\right)^{2}\sum_{L'}\int \rho'^2d\rho'\Gamma_{JL'}(r,\rho')
\chi^{J}_{L'L}(\rho',\rho).
\label{gamt}
\end{equation}
        \section{The Effects of Distortion}
For numerical calculations it is convenient to come back from expression
(\ref{res}) to more complex one similar to (\ref{fsec})
\begin{equation}
{{d^2{\sigma}}\over{dE^{\prime}d\Omega}} = {{M^2_{He}}\over{4{\pi}^2}}
{{p^\prime}\over{p}}\overline{\Gamma_{\pi Het}^\dagger\cdot
G^\ast\cdot \Im{m}\Pi\cdot G\cdot\Gamma_{\pi Het}}.
\label{ddsec}
\end{equation}
The product sign means integration over all coordinates in the configuration
space and the overline is the averaging and summing over spins of a $^3\!$He
and a triton.

In infinite matter (\ref{ddsec}) would be the only contribution to the
inclusive cross--section. In finite nucleus, however, the cut of a bare
pionic line gives nonzero contribution corresponding to coherent pion
production.
\begin{equation}
{{d^2{\sigma}_c}\over{dE^{\prime}d\Omega}} = -{{M^2_{He}}\over{4{\pi}^2}}
{{p^\prime}\over{p}}\overline{\Gamma_{\pi Het}^\dagger\cdot
G^\ast\cdot \Pi^\ast\cdot \Im{m}G^0\cdot\Pi
\cdot G\cdot\Gamma_{\pi Het}}.
\label{cohpi}
\end{equation}

In the plane wave approximation the ${\pi\,^3}{\!}\mbox{He{\,t}}$
vertex is
\begin{equation}
\Gamma_{\pi Het}({\bf r}) = \sqrt{2}({\bf\sigma}\cdot{\bf \tilde{q}})
F(q^2){f_{N}(q^2)\over\mu}\exp(\imath{\bf qr}),
\label{vert}
\end{equation}
where the effective momentum transfer $\tilde{{\bf q}}$ in lab. system is
$$\tilde{\bf q} = \sqrt{\frac{E'+M}{E+M}}{\bf p} -
\sqrt{\frac{E+M}{E'+M}}{\bf p'},$$ here $E$ is the total energy of $^3\!$He
and $M$ is its mass.
At first order in $\omega\over{E+M}$ it can be rewritten as
$$\tilde{\bf q} = {\bf q} - {1\over2}{\omega\over{E+M}}({\bf p}+{\bf p'})$$
The effective momentum transfer squared coincides with the four--momentum
transfer squared.
$F(q^2)$ is the ($^3\!$He,t) transition form factor.

The multipole expansion of the vertex looks as follows
\begin{equation}
\Gamma_{\pi Het}({\bf r}) = \sum_{JLM}\Gamma^{N}_{LJM}(r)t^{L}_{JM}({\bf n}),
\end{equation}
where $t^{L}_{JM}({\bf n})$ are the tensor operators analogous
to the (\ref{tens})
\begin{equation}
t^{L}_{JM}({\bf n})= ({\bf\sigma}\cdot{\bf Y}^{L}_{JM}({\bf n}))
  = \left[{\bf\sigma}\wedge Y_{LM}\right]_{JM}.
\end{equation}
For plane waves the radial vertex is
\begin{equation}
\Gamma^{0N}_{JLM}(r) = \sqrt{2}(\imath)^{L}j_{L}(qr)\left[{\bf \tilde{q}}
\wedge Y^{\ast}_{Lm}({\bf\hat{q}})\right]_{JM}{f_{N}(q^2)\over\mu}F(q^2)
\end{equation}

The distortion of the incoming and outgoing waves has been taken into
account via inelastic distortion factor \cite{dmit89}. With this factor
the $\pi ^3\!$Het vertex becomes
\begin{equation}
\Gamma_{\pi Het}({\bf r}) = \sqrt{2}[-\imath({\bf\sigma}\cdot{\bf \nabla})
- {1\over2}\frac{\omega}{E+M}({\bf\sigma}\cdot({\bf p}+{\bf p'}))]
{f_{N}(q^2)\over\mu}\exp(\imath{\bf qr})\exp(-{1\over2}\chi_{in}({\bf r}_
{\perp},{\bf q})).
\label{dvert}
\end{equation}
 The distortion factor $\exp(-{1\over2}\chi_{in}({\bf r}_{\perp},{\bf q}))$
 is \cite{dmit89}
\begin{eqnarray}
\nonumber
  \exp(-{1\over2}\chi_{in}({\bf r}_{\perp},{\bf q}))=\left(1-{1\over A}
 \bar{\gamma}T({\bf r}_{\perp})\right)^{\textstyle A-1}\cdot\int d^{3}
 s_{1}d^{3}s_{2}d^{3}s_{3}
 \exp(\imath{\bf qs_1}){\Psi}^{\ast}({\bf s}_{1},{\bf s}_{2},{\bf s}_{3})
 \cdot \\
\nonumber
\left(1-{1\over A}\bar{\gamma}T({\bf r}_\perp+{\bf s}_{2\perp}-{\bf s}_{
1\perp})\right)^{\textstyle A}\cdot
\left(1-{1\over A}\bar{\gamma}T({\bf r}_\perp+{\bf s}_{3\perp}-
 {\bf s}_{1\perp})\right)^{\textstyle A}\cdot \\
 \Psi({\bf s}_{1},{\bf s}_{2},{\bf s}_{3})\delta({\bf s}_1 +
 {\bf s}_2 + {\bf s}_3),\hspace{0.5in}
 \label{chi}
 \end{eqnarray}
where $T({\bf r}_{\perp})$ is the thickness function $$T({\bf r}_{\perp})
= \int_{-\infty}^{\infty}\rho({\bf r}_{\perp},z) dz,$$
and $\rho({\bf r}_{\perp},z)$ is the target density. The $$\bar{\gamma}
= -\imath{{2\pi}\over{p_{lab}}}f(0)$$ is
related to the elastic nucleon--nucleon scattering amplitude at given
energy per nucleon and $\Psi({\bf s}_{1},{\bf s}_{2},{\bf s}_{3})$ is
the wave function of the $^3\!$He or the triton depending on the internal
coordinates ${\bf s}$. The value of $\bar{\gamma}$ used in calculations
is $\bar{\gamma} = (2.1 -\imath0.26) fm^2$ \cite{wall81}.

Two features of the distortion factor (\ref{chi}) should be mentioned.
First, the $^3\!$Het form factor can not be separated from the effects of
distortion. Second, since the vertex (\ref{dvert}) has a gradient coupling
and the distortion factor (\ref{chi}) depends on the transversal coordinates,
some transversal components arise in the reaction amplitude even if it was
before pure longitudinal amplitude.

The multipole expansion of the distorted vertex (\ref{dvert}) can be obtained
directly by multiplying it on $t^{\dagger L}_{JM}({\bf n})$, taking trace over
spin matrices, and integrating over angles of the unit radius vector ${\bf n}$.
\begin{equation}
\Gamma^{N}_{LJM}(r) = {1\over 2}Tr\int d{\bf n}{\,}\left(t^{\dagger L}_{JM}
({\bf n})\cdot \Gamma_{\pi Het}({\bf r})\right).
\end{equation}

The separate multipoles contribute independently into cross--section
(\ref{ddsec}), which becomes
\begin{equation}
{{d^2{\sigma}}\over{dE^{\prime}d\Omega}} = {{M^2_{He}}\over{4{\pi}^3}}
{{p^\prime}\over{p}}\sum_{LJM}{\Gamma_{LJM}^{\ast N}\cdot
G^\ast_{L}\cdot \Im{m}\Pi_{L}\cdot G_{L}\cdot\Gamma_{LJM}^{N}}.
\label{dsecL}
\end{equation}
Similar expansion exist for the coherent pion contribution
(\ref{cohpi}).

For numerical calculation it is convenient to define the function
$$w_{LJM}(r) = \int_{0}^{\infty}r'^2dr'\,G_{L}(r,r')\Gamma_{LJM}^{N}(r'),
\label{int}$$
which is the pion field at the reaction point generated by the source
$\Gamma_{LJM}^{N}(r')$. The integration in (\ref{int}) is not well
defined numerically since the integrand is oscillating function that is
not decreasing at infinity. The indirect integration has been done in
the following way.
The function $w_{LJM}(r)$ satisfies  the integro--differential equation
\begin{equation}
\int_{0}^{\infty}r'^{2}dr'\,G^{-1}_{L}(r,r')w_{LJM}(r') = \Gamma^{N}_{LJM}(r),
\label{gout}
\end{equation}
which was solved numerically using the condition for Feynman propagator
$G_{L}(r,r')= G_{L}^{(+)}(r,r')$ at positive energy.
Since $G_{L}^{(+)}(r,r')$ has an outgoing wave at infinity it fixes the
solution of the equation (\ref{gout}). The cross--section expressed in terms
of $w_{LJM}(r)$ is
\begin{equation}
{{d^2{\sigma}}\over{dE^{\prime}d\Omega}} = {{M^2_{He}}\over{4{\pi}^3}}
{{p^\prime}\over{p}}\sum_{LJM}{w_{LJM}^{\ast}\cdot
 (\Im{m}\Pi_{L} - \Pi_{L}^{\ast}\cdot\Im{m}G^0\cdot\Pi_{L})\cdot w_{LJM}}.
\label{dsecw}
\end{equation}
In the expression (\ref{dsecw}) the integration over coordinates goes
effectively in a finite range, inside the target nucleus.
\section{The Triton Spectra for $^{12}\!C(\,^3\!He,t)$ Reaction at 2 GeV}
\label{cross}
\subsection{Parameters of the single--particle potentials.}
The nucleon single--particle potential used for  the
wave functions of the bound nucleons has been taken in the standard
Woods--Saxon form.
$$ U(r) = V_{0}\cdot f(r) + V_{LS}{\lambda_{\pi}^2\over{r}}{df(r)\over{dr}}
 ({\bf \sigma}\cdot{\bf l}) + V_{C}(r),$$ where $f(r) = \frac{1}{1 +
 \exp({{r-R}\over{a}})}$,
 $\lambda_{\pi}$
 is the pion Compton wavelength, and $V_{C}(r)$ is the Coulomb potential for
 protons that was taken as the potential of a uniformly charged sphere.
 The parameters of the potential are listed in the Table 1. The response
 function (\ref{fresp}) were found not very sensitive to the parameters
 of the nucleon potential.

 The situation is, however, different for the optical $\Delta$ -- nucleus
 potential that has been taken in similar Woods--Saxon form.
$$ U_{\Delta}(r) = (V_{\Delta}+ \imath W_{\Delta})\cdot f(r) +
(V_{\Delta LS}+ \imath W_{\Delta LS}){\lambda_{\pi}^2
\over{r}}{df(r)\over{dr}}({\bf s}_{\Delta}\cdot{\bf l}) + V_{\Delta C}(r),$$
where $s_{\Delta}$ are the spin $3/2$ matrices. Fig. 2 shows two triton
spectra for the $\Delta$--h contribution demonstrating sensitivity to
the real part of the $\Delta$--nucleus optical potential. In the absence
of the $\Delta$--nucleus interaction the peak position coincides with
the one in the reaction on free proton. The attraction produced by the
real part of the potential in the final state increases cross-section
and shifts the peak position on 15-20 MeV down.
The parameters of the optical
potential are listed in Table 2. They were taken mainly from \cite{lee82},
\cite{koch84}, and \cite{udaga90}.  The radius $R$ and the diffuseness $a$
were kept the same as for the nucleons.
\subsection{Medium Effects of Pion Renormalization}
 The main feature of the pionic self--energy near the resonance is its
large imaginary part. It is instructive to study separately
the effects of real and imaginary parts of the self--energy on the
triton spectra. Fig.3 shows three spectra where either real or imaginary
part of the self--energy were accounted in comparison to the quasifree
case. The real part of the self--energy is attractive and it brings more
strength to the $\Delta$-hole peak. In analogy with particle--hole
excitations one can say the real part of the pionic self--energy
makes the $\Delta$-hole peak more collective. The peak position, however,
changes not much. The imaginary part, takes all this
collectivity back decreasing cross--section due to incoherent $\Delta$
decay so, when both parts are taken into account the
cross--section appeared close to its quasifree magnitude as it is shown
in Fig.3 and Fig.4.

In Fig.3 one can see that the imaginary part of the self--energy
produces also some shift in the peak position. The origin of the shift is,
however, different from the shift due to the real part. To understand
its origin let us return to nuclear matter and compare two expressions
for the cross-sections with and without renormalization of the pion
propagator. Without renormalization the cross--section is proportional
to
$${\Im{m}\Pi(\omega,{\bf q})\over{(q^2 - \mu^2)^2}},$$
while in the other case it is proportional to
$${\Im{m}\Pi(\omega,{\bf q})\over{(q^2 - \mu^2)^2 + (\Im{m}\Pi(\omega,{\bf
q}))^2}}.$$
For simplicity the real part is omitted in this expression. The first
case corresponds to quasifree mechanism and the peak position is at
the same place as in the reaction on a proton. In the second case the
cross-section starts to grow at the same threshold but at the resonance
where $\Im{m}\Pi(\omega,{\bf q}) > {(q^2 - \mu^2)}$ we have the
cross--section proportional to $${1\over{\Im{m}\Pi(\omega,{\bf q})}}.$$
Thus, instead of maximum the cross--section appeared to be small at this
energy. Since the
cross--section is growing from threshold one can immediately conclude that
the maximum of the cross-section will be below the resonance position.
Fig.5 demonstrates this effect for the $L=0$ multipole in the case of
finite nucleus. This effect is not so much pronounced for higher
multipoles that determine the magnitude of the cross--section. The
higher multipoles undergo smaller medium effects since
they are peaked at nuclear surface where the density is small. Therefore,
the overall shift of the peak position is smaller than for $L=0$.
The imaginary part of the optical $\Delta$ - nucleus potential makes
this effect stronger as it is shown in Fig.6.
\subsection{Contribution of separate multipoles.}
The contribution of separate multipoles to the triton spectrum is shown
in the Fig.7. The contribution of the low multipoles $L=0$ and $L=1$
is almost negligible due to strong absorption of the incoming and
outgoing ions.
The main contribution comes from the multipoles between $L=2$ to $L=6$
although higher multipoles, at least up to $L=10$, have to be cosidered
at higher excitation energy.

Another feature clearly seen in the Fig.7 is rather wide spreading of the
different multipole contributions. The $L=2$ contribution is
most sensitive to the medium effects shifting down the transition strength.
It has the largest downward shift in the peak position. The absorption of
the $^3\!$He and t is smaller for $L=2$ compared to $L=0$ resulting in
sizeable contribution
to the cross--section. Higher multipoles have smaller medium effects and
their peak positions are at more and more high excitation energies. This
produce large displacement width of summed triton spectrum.
\subsection{Coherent pion production.}
Fig.8 shows the final triton spectrum toghether with separate
contributions of the $\Delta$--hole excitations and the coherent pion
production. The process of coherent pion production gives sizeable
contribution to the inclusive triton spectrum. The maximum of the
cross--section is at about 240 MeV excitation energy and it also
contributes to the shift of the inclusive peak. The coherent pion
production is absent in infinite nuclear matter, therefore one should
expect decrease of the relative yield of coherent pions for heavier
nuclei.
\section{Conclusions}
For ($^3\!$He,t) reaction in the $\Delta$-region strong deviations from
impulse approximations were demonstrated. The deviations come from the
medium effects of renormalization of the pion propagator in the OPE mechanism
of the elementary charge--exchange reaction. The medium effects change
both the
peak position and its height. Several effects, besides pion
renormalization, contribute to the shift of
peak position including $\Delta$--nucleus optical potential,
and coherent pion production. The possible effects of collectivity in the
$\Delta$--hole excitations are strongly suppressed by incoherent
$\Delta$ decay. The effects of virtual pion propagation manifest itself
only in the process of coherent pion production.

The finite size of a target nucleus produces together with the medium
effects large spreading of the observed peak. The absorption in initial and
final states strongly supresses the lowest multipols of the angular
momentum transfer. The Glauber approach to the  distortion in initial and
final states gives reasonable description both the size of the medium effects
and the absolute value of the cross--section.

\acknowledgments{
During all stages of this work the discussions with Carl Gaarde and Thomas
Sams were very stimulating and helpful. I would like to thank
V.G.Zelevinsky and V.B.Telitsin as well for discussions of  different
aspects of the problem.}

\newpage
\begin{table}
\begin{tabular}{ccccccc}
   &$V_{0}$(MeV)&$V_{LS}$(MeV)&$R$(fm)&$R_{LS}$(fm)&$a$(fm)&$a_{LS}$(fm) \\
\hline
 p & 57         &12&$1.25\cdot A^{1/3}$&$1.25\cdot A^{1/3}$&0.53&0.53 \\
 n & 57        &12&$1.25\cdot A^{1/3}$&$1.25\cdot A^{1/3}$&0.53&0.53 \\
\end{tabular}
\caption{Parameters of the single--particle nucleon potential.}
\end{table}
\begin{table}
\begin{tabular}{cccc}
$V_{\Delta}$(MeV)&$W_{\Delta}$(MeV)&$V_{LS\Delta}$(MeV)&$W_{LS\Delta}$(MeV) \\
\hline
35 & 40 & 5 & 0 \\
\end{tabular}
\caption{Parameters of the $\Delta$- nucleus optical potential. The radius
         and diffuseness are the same as for the nucleons.}
\end{table}
\newpage

\figure{Direct OPE graph for the $\Delta$ production.\label{fig1}}
\figure{ Quasifree $\Delta$ production. Dashed line - no $\Delta$-nucleus
potential. Solid line - $V_{\Delta} = -35$ MeV\label{fig2}}
\figure{Effects of real and imaginary parts of the pion self-energy on
the $\Delta$-h part of the triton spectrum.
Dotted line -- no medium effects for pion. Dashed line -- effect of the
real part of the self-energy. Solid line -- effect of the imaginary part
of the pion self-energy.\label{fig3}}
\figure{Dashed line -- effect of the real part. Dotted line -- effect of
the imaginary part. Solid line -- full pion self-energy included.\label{fig4}}
\figure{The shift of the peak position due to imaginary part of the pion
self-energy for
$L=0$ $\Delta$-h multipole.\label{fig5}}
\figure{Influence of the imaginary part of the $\Delta$-nucleus optical
potential. Dashed line $W_{\Delta}=0$. Solid line $W_{\Delta}= 40$
MeV.\label{fig6}}
\figure{Contribution of separate $\Delta$-h multipoles.\label{fig7}}
\figure{Different contributions to the triton spectrum. Dashed line --
$\Delta$-h contribution. Dotted line -- coherent pion production. Solid
line -- full spectrum. \label{fig8}}

\newpage
\setlength{\unitlength}{1mm}
\begin{picture}(50,30)(-60,0)
\put(0,0){\vector(1,0){15}}
\put(0,15){\vector(1,0){15}}
\put(15,15){\vector(1,0){15}}
\multiput(15,0)(0,2){8}{\line(0,1){1}}
\thicklines
\put(15,0){\vector(1,0){15}}
\put(2,18){p}
\put(28,18){n}
\put(2,-3){p}
\put(28,-4){$\Delta^{++}$}
\put(12,-10){Fig.1}
\end{picture}

\begin{references}
\bibitem{elle83}C. Ellegaard et al., Phys. Rev. Lett. {\bf 50} (1983) 1745.
\bibitem{able84}V.G. Ableev et al., JETP Lett. {\bf 40} (1984) 763.
\bibitem{cont86}D. Contardo et al.,Phys. Lett. {\bf 168B} (1986) 331.
\bibitem{gaard91}C. Gaarde, Annu. Rev. Nuc. Part. Sci.,{\bf 41} (1991) 187.
\bibitem{chanf84}G. Chanfray and M.Ericson, Phys. Lett. {\bf 141B} (1984)
 163.
\bibitem{dmit85}V.F. Dmitriev and Toru Suzuki, Nucl. Phys. {\bf A438}
 (1985) 697. V.F. Dmitriev, Yad. Fiz. {\bf 46} (1987) 770.
\bibitem{udaga90}S.W. Hong, F. Osterfeld and T. Udagawa, Phys. Lett.
 {\bf 245B} (1990) 1.
\bibitem{delo91}J. Delorm and P.A.M. Guichon, Phys. Lett. {\bf 263B}
 (1991) 157.
\bibitem{oset89}E. Oset, E. Shiino and H. Toki, Phys. Lett. {\bf 224B}
(1989) 249.
\bibitem{esbe85}H. Esbensen and T.-S.H. Lee, Phys. Rev. {\bf C32} (1985)
 1966.
\bibitem{dmit89}V.F. Dmitriev, Phys. Lett. {\bf 226B} (1989) 219.
\bibitem{ferr61}E. Ferrari, F. Selleri, Phys.Rev. Lett. {\bf 7} (1961)
 387.
\bibitem{wolf69}G. Wolf, Phys. Rev. {\bf 182} (1969) 1538.
\bibitem{dmit86} V.F. Dmitriev, O.P. Sushkov and C. Gaarde, Nucl.Phys.
 {\bf A459} (1986) 503.
\bibitem{verv79}B.J. Vervest, Phys.Lett. {\bf 83B} (1979) 161.
\bibitem{karm73}V.A. Karmanov et al., Sov.Journ.Nucl.Phys. {\bf 18}
 (1973) 1133.
\bibitem{desg92}P. Desgrolard, J. Delorme and C. Gignoux, preprint
 Inst. de Phys. Nucl. de Lyon, LYCEN 9205 (1992).
\bibitem{olss75}M.G. Olsson, E.T.Osypowsky, Nucl.Phys. {bf B101} (1975) 136.
\bibitem{erer66}M. Ericson and T.E.O. Ericson, Ann. Phys. {\bf 36} (1966) 323.
\bibitem{lee82} T.-S. H. Lee and K. Ohta, Phys.Rev. C{\bf 25} (1982) 3043.
\bibitem{koch84}J. H. Koch, E. J. Moniz and N. Ohtsuka, Ann. of Phys.
                {\bf 154} (1984) 99.
\bibitem{wall81}Stephen J. Wallace, in Advances in Nuclear Physics,
                Plenum Press, New York, Vol. 12, 1981.
\end{references}
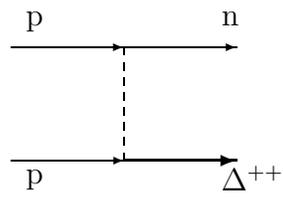
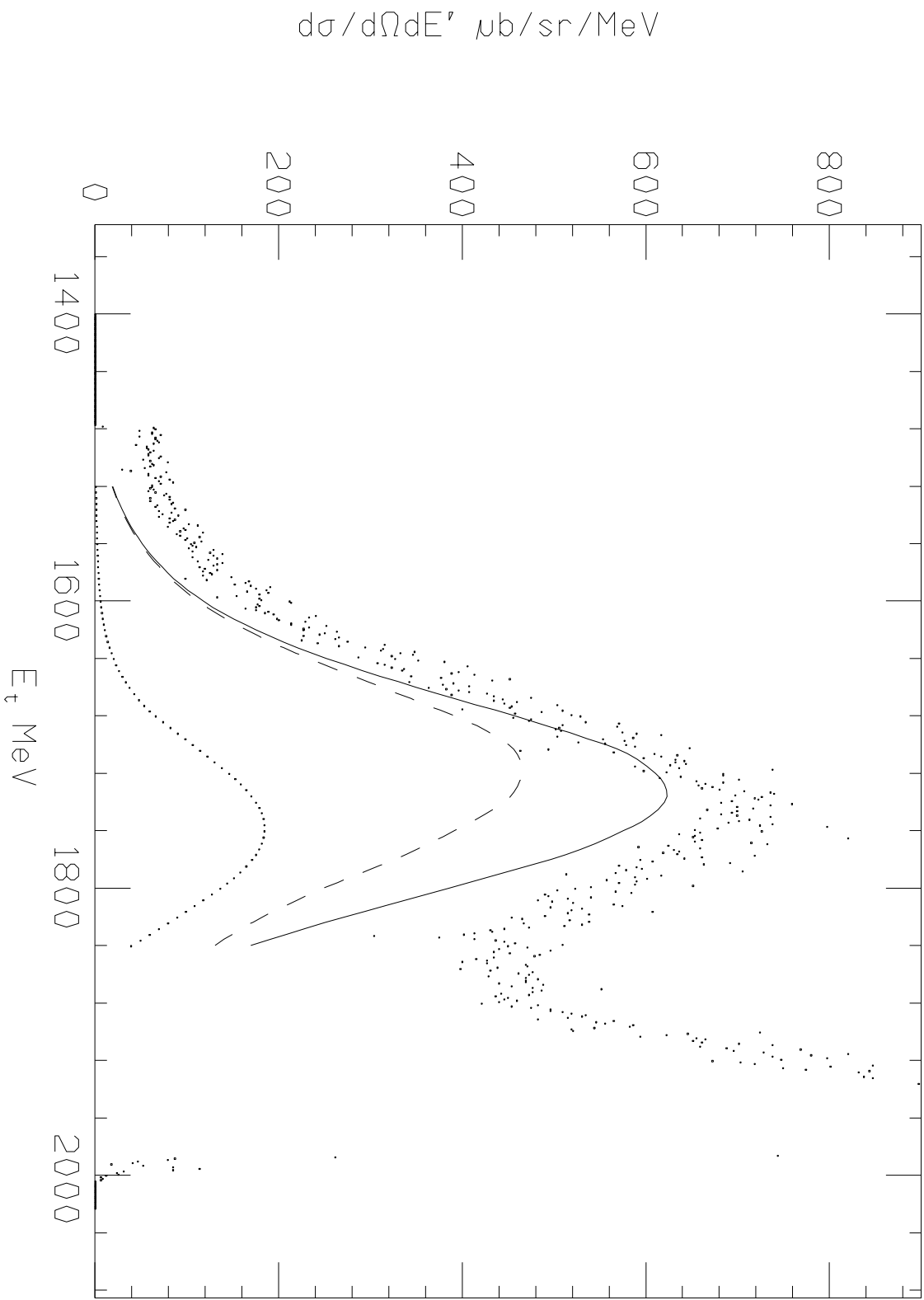
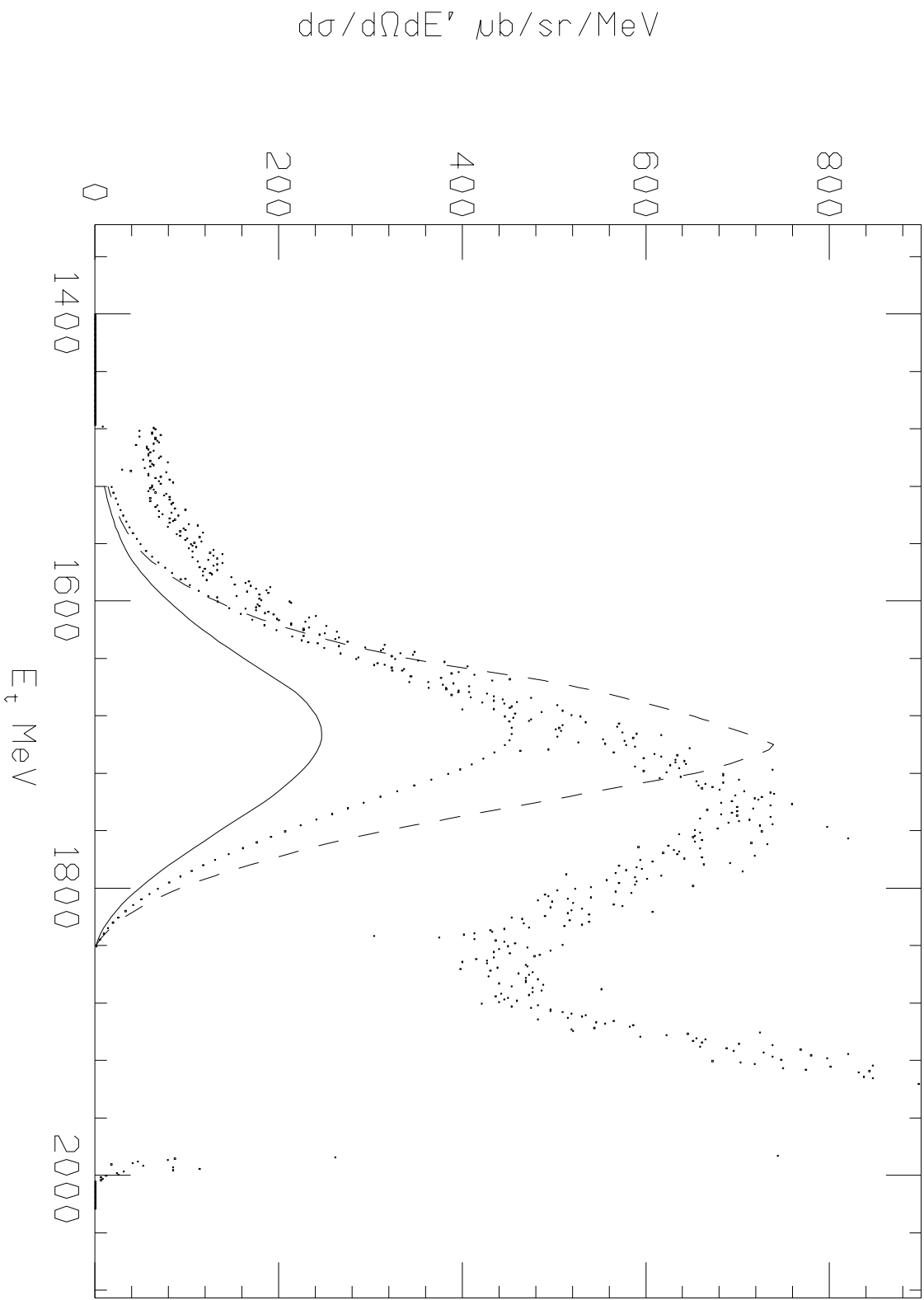
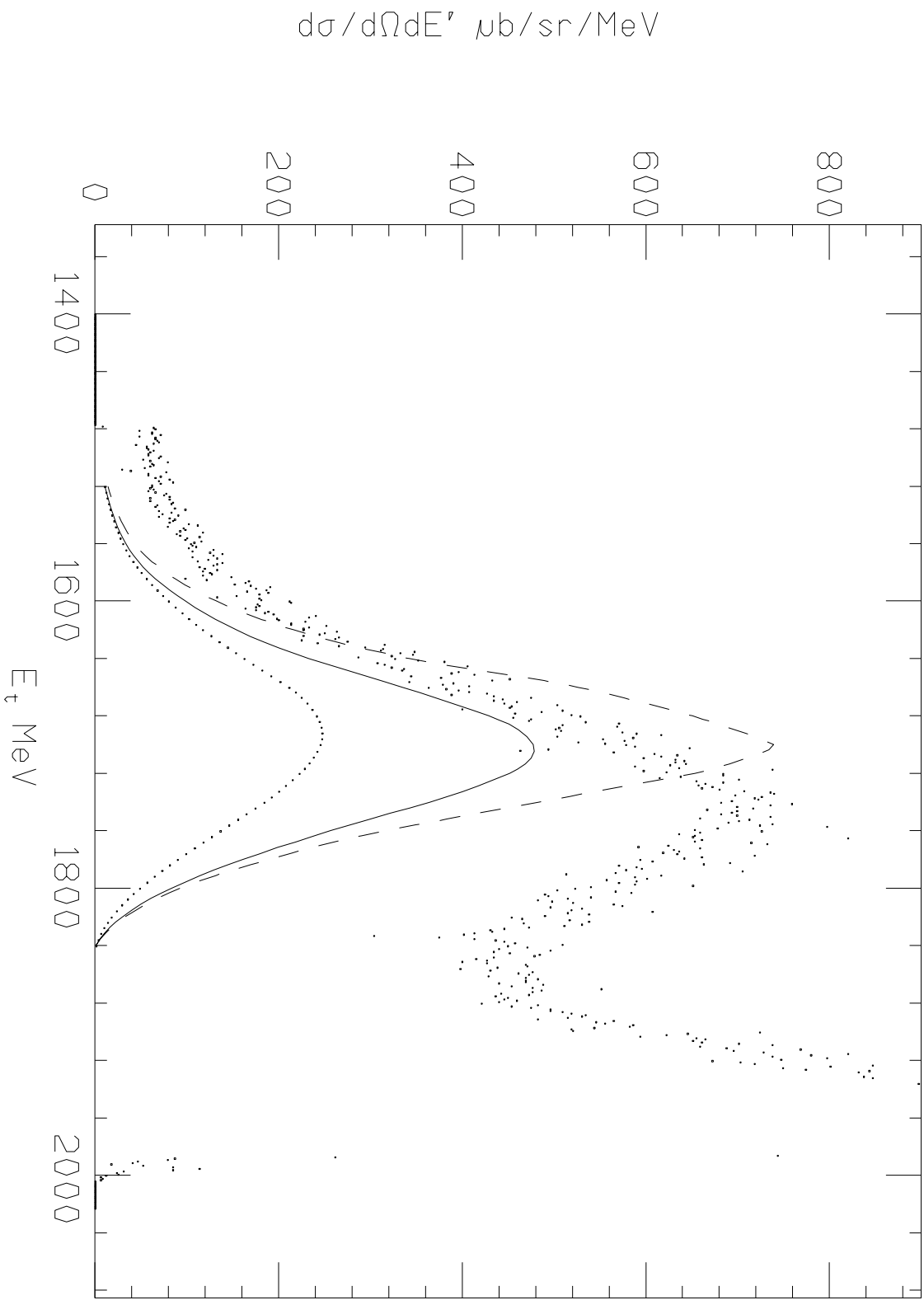
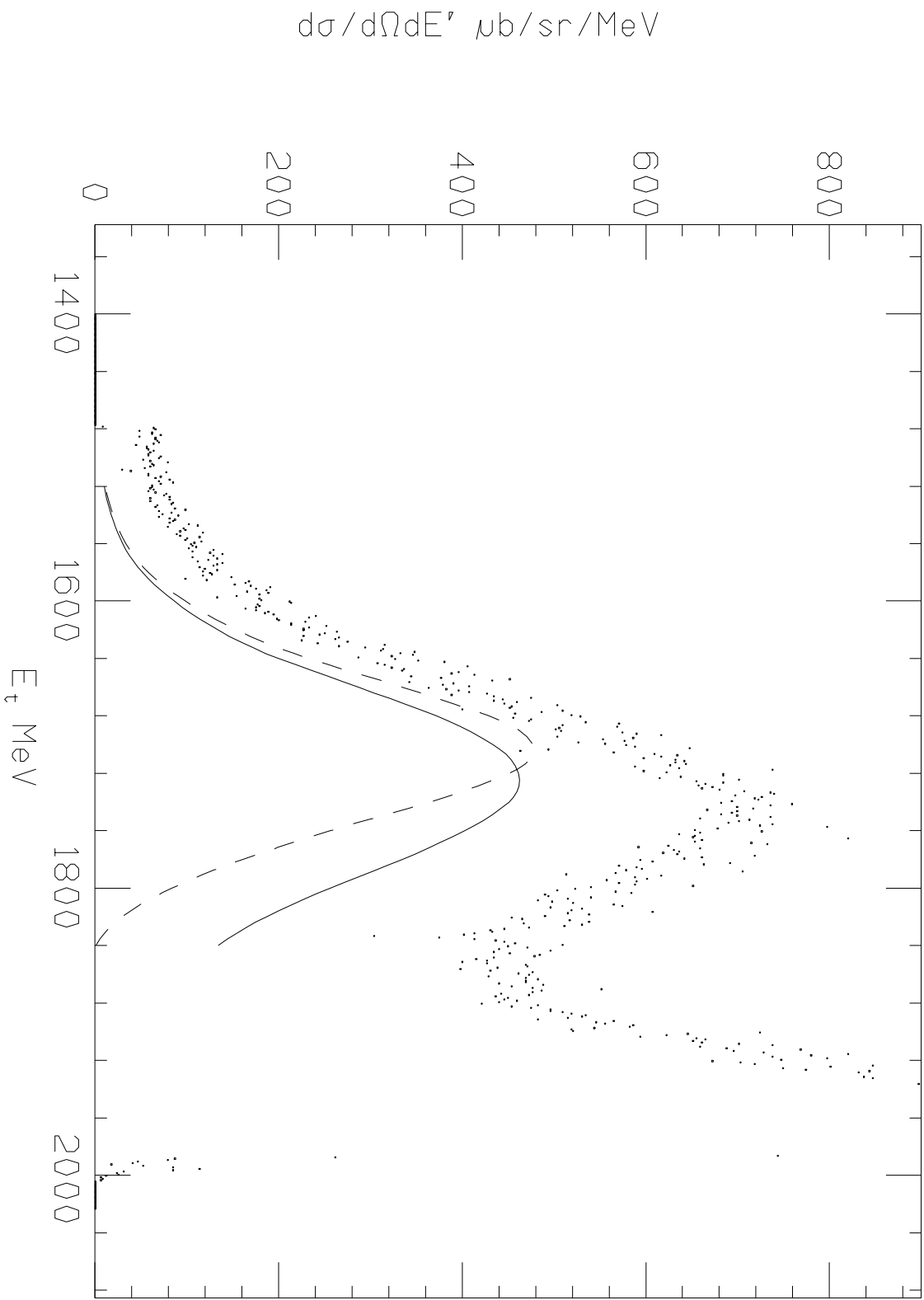
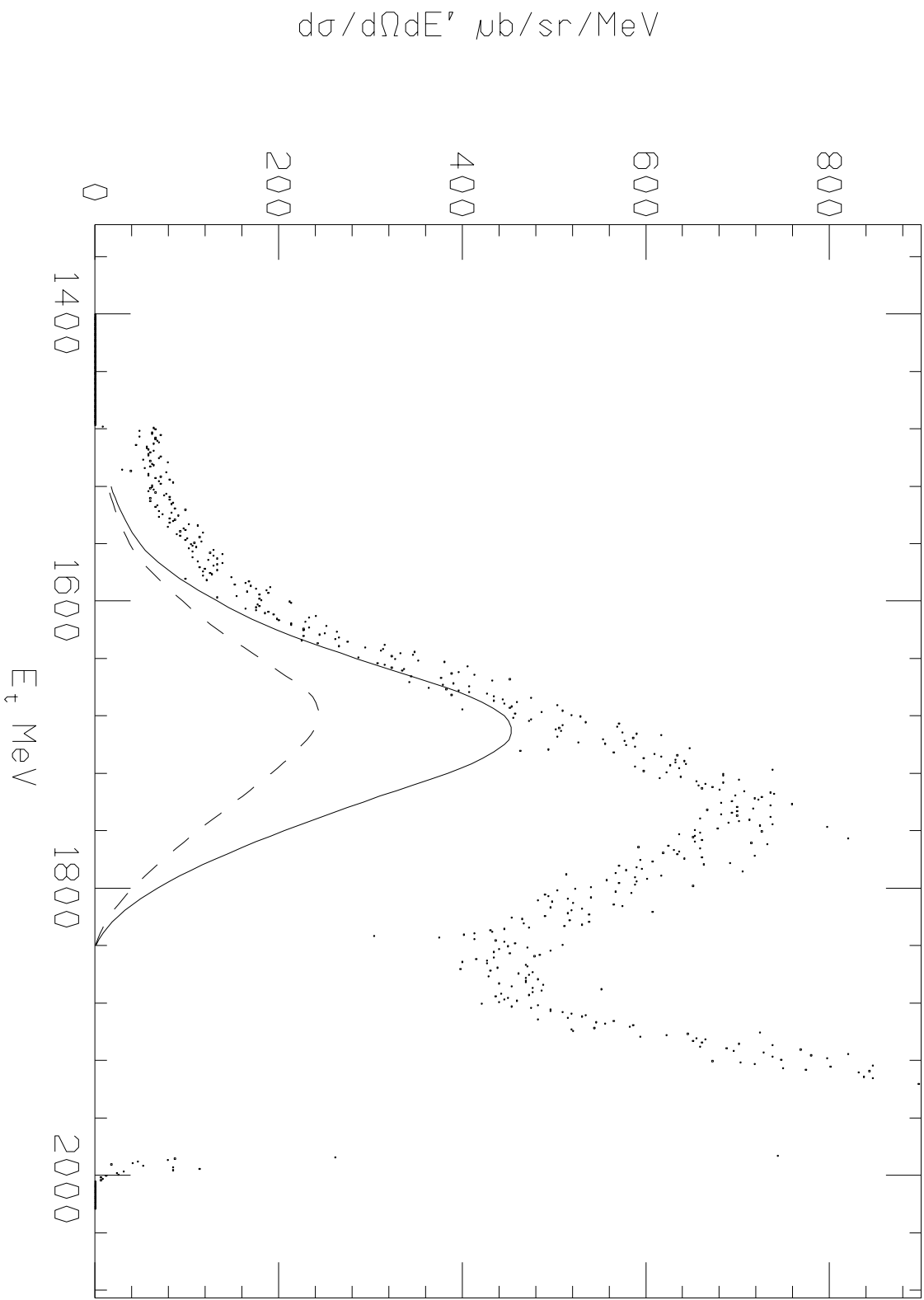
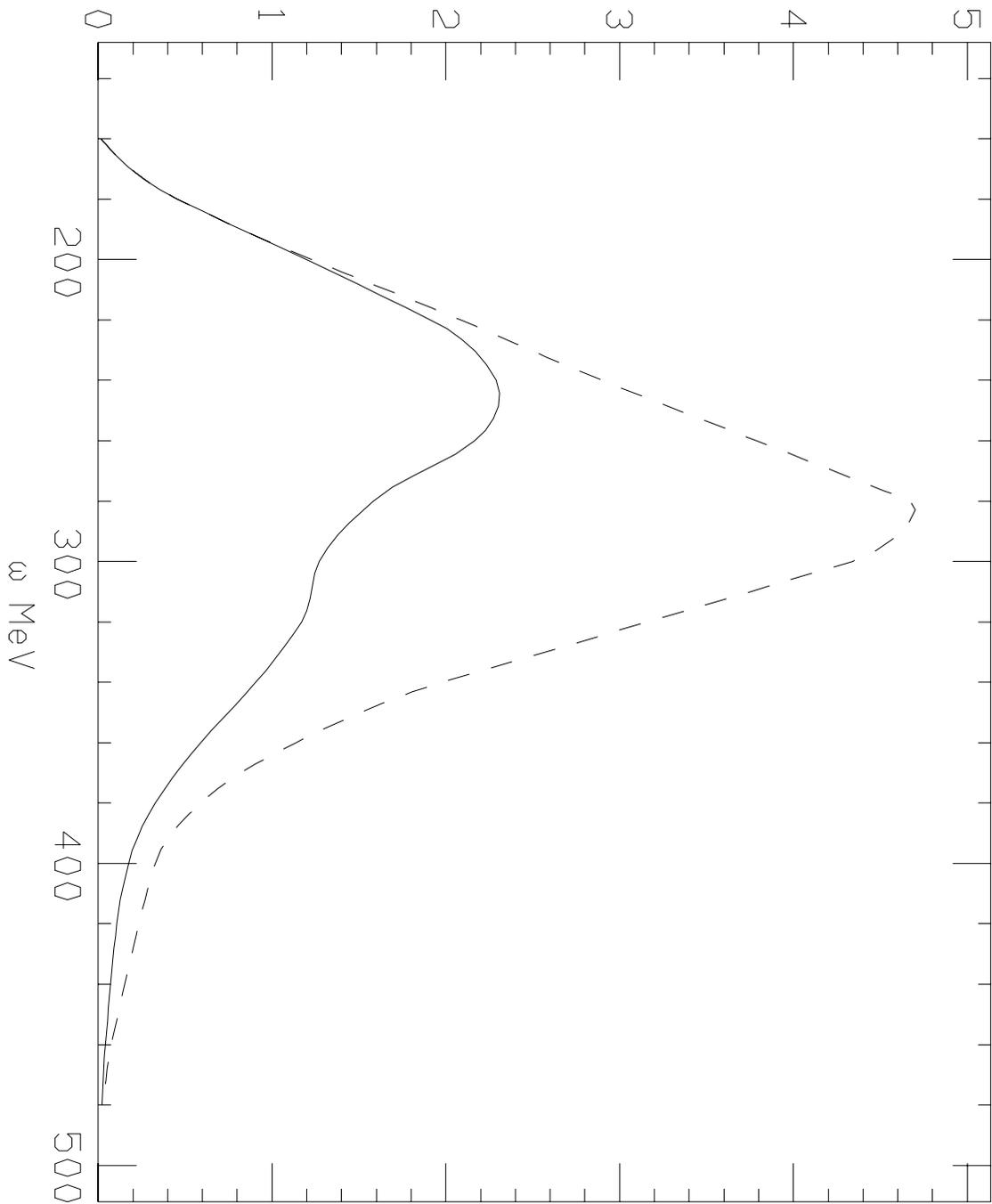
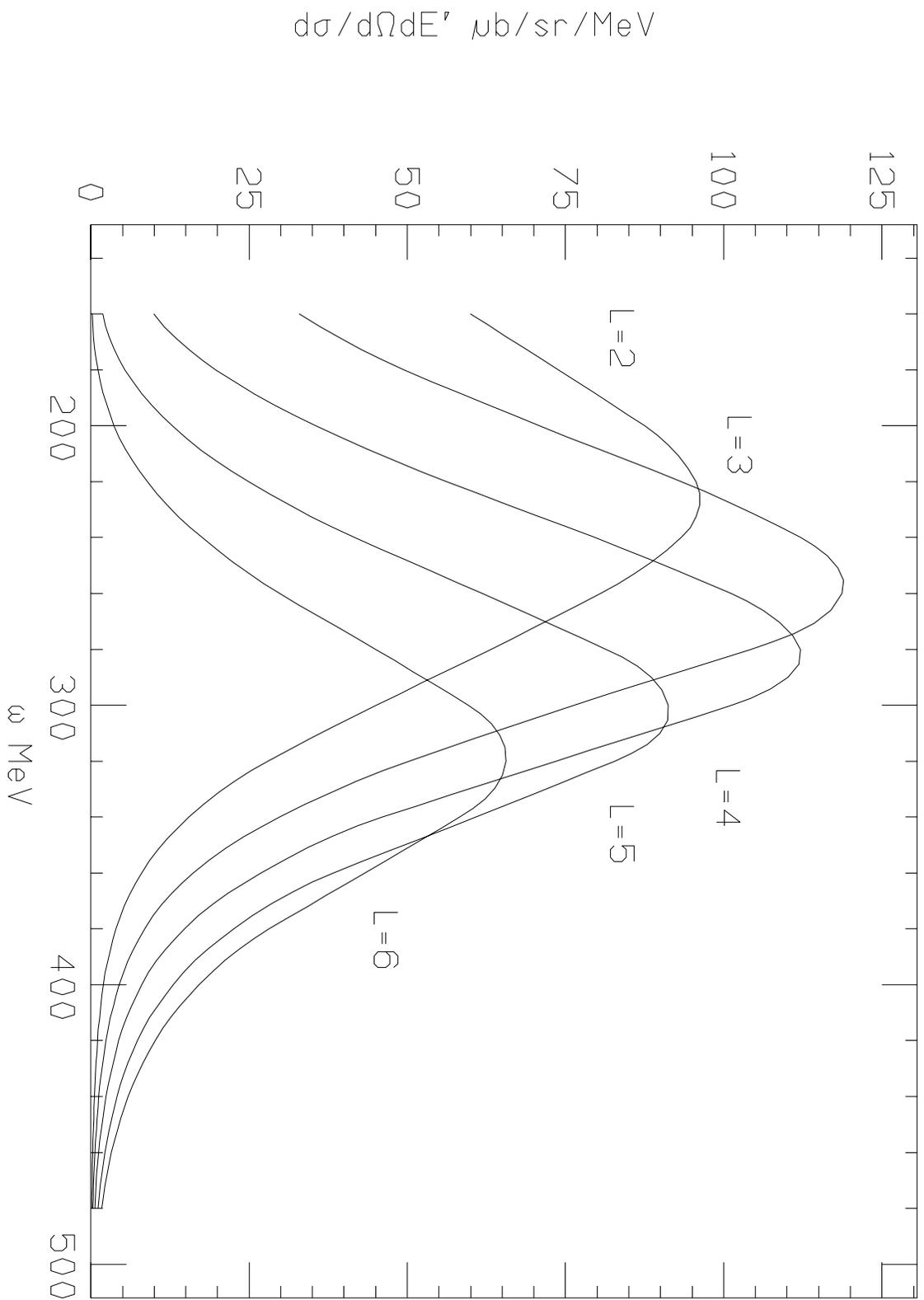
\end{document}